\documentstyle[12pt]{article} \topmargin=-15mm \textwidth=155mm
\def\pa{\partial}
\def\al{\alpha}
\def\ga{\gamma}
\def\dl{\delta}
\def\be{\beta}
\def\nb{\nabla}
\def\sg{\sigma}
\def\ck{\check}
\def\la{\lambda}

\title{\bf New Approach to $N$-body Relativistic \\Quantum Mechanics}
\author{Ying-Qiu Gu\footnote {email: yqgu@luody.com.cn}}
\date{\small Department of Mathematics, Fudan University Shanghai, 200433, China}
\begin{document}
\maketitle

\begin{abstract}
In this paper, we propose a new approach to the relativistic
quantum mechanics for many-body, which is a self-consistent system
constructed by juxtaposed but mutually coupled nonlinear Dirac's
equations. The classical approximation of this approach provides
the exact Newtonian dynamics for many-body, and the
nonrelativistic approximation gives the complete Schr\"odinger
equation for many-body.


\end{abstract}

\section{Introduction}
\setcounter{equation}{0}

The nonrelativistic quantum mechanics for many body is described
by the Schr\"odinger equation, which has successful and fruitful
applications in quantum chemistry, quantum statistics, condensed
matter physics, etc. The wonderful properties of this equation
leads to many refined mathematical methods to solve the
eigenvalues and eigenstates\cite{1}.

But the corresponding relativistic theory is an unsettled problem.
There are several approaches for this problem. Early in 1937, Fock
proposed a 4+1 dimensional covariant Schr\"odinger-like equation
with a scalar time\cite{2}. In 1941, St\"uckelberg also provided a
4+1 dimensional covariant equation but without specific
mass\cite{3}. The equation was improved and generalized to the
case with multi proper time\cite{4,5,6,7,8,9,10}.

A manifestly covariant approach derived from quantum field theory
was introduced by Bethe and Salpeter\cite{11}, and developed in
\cite{12,13}. But it is too complicated for practical calculation
for more than 2 particles. A alternative version is the mass shell
constraints approach given in \cite{14,15,16,17,18}, which has to
treat multi relative times. This approach also is effective for
two or three bodies problem.  In \cite{19,20}, After analyzing the
former approaches, the authors proposed a new method to define a
uniform time for n-particle with the same mass. They applied
generalized Foldy-Wouthuysen transformation to the positive energy
part of the Hamiltonian, and computed the charmonium and bottonium
eigenstates.

Although the authors of \cite{19} mentioned that to treat
relativistic quantum mechanics for many body problem as a field
theory was essentially abandoned. However all matters are
4-dimensional existence, whose intrinsic properties should be
naturally described by $3+1$ dimensional field variables. In this
paper, we propose a new approach in the form of nonlinear coupled
field theory, which assign a different spinor field $\phi_k$
instead of coordinate $x^\mu_k$ to each particle. In this version
every thing can be well defined and conveniently for mathematical
treatment. We derive the Newtonian mechanics for $N$-electron via
clear definition and deduction, and derive the corresponding
$N$-body Schr\"odinger equation under nonrelativistic
approximation. The results show that it seems to be a meaningful
model.

\section{The Fundamental Equations}
\setcounter{equation}{0}

Denote the Minkowski metric by $\eta_{\mu\nu}={\rm
diag}[1,-1,-1,-1]$, Pauli matrices by
\begin{equation}
 {\vec\sg}=(\sg^{j})= \left \{\pmatrix{
 0 & 1 \cr 1 & 0},\pmatrix{
 0 & -i \cr i & 0},\pmatrix{
 1 & 0 \cr 0 & -1}
 \right\}.\label{1.1}\end{equation}
Define $4\times4$ Hermitian matrices as follows
\begin{equation}\al^\mu=\left\{\left ( \begin{array}{ll} I & ~0 \\
0 & I \end{array} \right),\left (\begin{array}{ll} 0 & \vec\sg \\
\vec\sg & 0 \end{array}
\right)\right\},~ \ga =\left ( \begin{array}{ll} I & ~0 \\
0 & -I \end{array} \right),
~ \be=\left (\begin{array}{ll} 0 & -iI \\
iI & ~~0 \end{array} \right).\label{mtr}
\end{equation}
For $N$ electrons $e_1, e_2,\cdots,e_N$ moving in the external
potential $V^{\mu}$ generated by external charge $\rho^\mu$, we
find that the following Lagrangian give a good description for the
system of $N$-electron,
\begin{equation}
{\cal L} =\sum^{n}_{k=1}\phi_k^+[\al^\mu (\hbar i\pa_\mu-eA_\mu)
-\mu
\ga]\phi_k-\rho_{\mu}A^{\mu}+\frac{1}{2}\pa_{\mu}A_{\nu}\pa^{\mu}A^{\nu}+F,
\label{2.9}\end{equation} where $\mu>0$ is a constant mass, which
takes one value for the same kind particles. $F$ is the nonlinear
coupling term, we take the following form as example for
calculation
\begin{equation}F=\frac{1}{2}w(Q_{\mu}Q^{\mu} -\sum^N_{k=1}\ck\be_k^2),\quad w>0,
\label{20}\end{equation} where $Q^{\mu}$ is the total current
\begin{equation}
 Q^{\mu}=\sum^N_{k=1}q^{\mu}_k,\quad q_k^\mu=\phi^{+}_k\al^\mu\phi_k, \quad
\ck\be_k=\phi^{+}_{k}\be\phi_k, \label{2.3}
\end{equation}
In this paper, we adopt the Hermitian matrices (\ref{mtr}) instead
of Dirac matrices $\ga^\mu$, because this form is more convenient
for calculation.

The variation of (\ref{2.9}) with respect to $\phi_k$ gives the
dynamic equation for $e_k$
\begin{equation}
\al^\mu(\hbar i\pa_\mu-eA_\mu+wQ_\mu)\phi_k=(\mu \ga +w
\ck\be_k\be) \phi_k, \label{2.2}
\end{equation}
or in the Hamiltonian form as usual
\begin{equation}
\hbar i \pa_t\phi_k=\hat H_k\phi_k,~~(k=1,2,\cdots,N)\label{ham}
\end{equation}
where the nonlinear Hamiltonian operator $\hat H_k$ is defined by
\begin{equation}
\hat H_k\equiv \vec\al\cdot[-\hbar i \nb -e\vec A+w(\vec Q-\vec
q_k)]+e A_0-w(Q_0-q_{k0})+(\mu -w\ck\ga_k)\ga.\label{21}
\end{equation}
with $\ck\ga_k\equiv \phi_k^+\ga\phi_k$. In (\ref{21}) we have
used the Pauli-Fierz identity
$q_{k\mu}q^{\mu}_k=\ck\be^2_k+\ck\ga_k^2$ \cite{21}. In this paper
we denote $\vec A=(A^1,A^2,A^3)$ to be the spatial part of a
contravariant vector $A^\mu$.

For each bispinor $\phi_k$, it is easy to check that the current
conservation law holds $ \pa_{\mu}q^{\mu}_k=0$, so we can take the
normalizing condition as
\begin{equation}
 \int_{R^3}q^0_k d^3x=\int_{R^3}|\phi_k|^2d^3x=1, \quad (\forall k).
\label{2.5}\end{equation} The electromagnetic theory is the
following Maxwell equation
\begin{equation}
  \left \{\begin{array}{ll}
   \pa_\mu q^{\mu}=\pa_\mu A^{\mu}=0, & \pa^\al\pa_\al A^{\mu}= eQ^{
\mu}+\rho^\mu,\\
   \vec{E}=-\nb A^0-\pa_0\vec{A}, & \vec{B}=\nb \times \vec
{A},\\
   \nb\cdot\vec{E}=eQ^0+\rho^0, & \nb\times\vec{E}=-\pa_0\vec{B},\\
   \nb\cdot\vec{B}=0, & \nb\times \vec{B}=\pa_0\vec{E}+e\vec{Q}+\vec \rho,
  \end{array}\right.\label{2.8}\end{equation}

(\ref{ham}) and (\ref{2.8}) is our starting point. This system has
many important properties. The numerical results reveal
\cite{22,23,24,25,26,27,28,29}:

(P1) All eigenstates have only positive mass spectra.

(P2) For normal eigenstate, its mean diameter is about $10^3$
Compton wave length, and its total energy is almost equal to $\mu
c^2$.

(P3) The energy contributed by the nonlinear term and its own
electromagnetic field is less than $1eV$ and very stable.

(P4) The abnormal magneton exists for nonlinear Dirac's equation
with its own electromagnetic field.

(P5) The different spinor can not take the same eigenstate.

In what follows, we define some classical concepts such as
coordinate, speed, mass etc. for an electron $\phi_k$
respectively, then derive their dynamic equation

{\bf Definition 1.} {\em The coordinate $X(t)_k$ and
speed $v_k$ of $k$-th electron $e_k$ are defined respectively by
\begin{equation}
{\vec X}_k(t) \equiv \int_{R^3}\vec{x}|\phi_k|^2 d^3
x=\int_{R^3}\vec x q_k^0d^3x,\quad {\vec v}_k\equiv\frac d
{dt}\vec X_k. \label{3.1}\end{equation}}

{\bf Definition 2.} {\em Define the 4-dimensional momentum
$p^\mu_k$ and energy $K_k$ of $e_k$  respectively by
\begin{equation}\left\{\begin{array}{l}
p_k^\mu\equiv\int_{R^3}\phi^{+}_{k}[\hbar
i\pa^\mu-eA^\mu+w(Q^\mu-q_k^\mu)]\phi_kd^3x. \\
K_k \equiv p^0_k+W_k,~~W_k\equiv \frac 1 2 w\int\ck\ga_k^2
d^3x.\end{array}\right.
 \label{3.3}\end{equation}}

{\bf Lemma 1.} {\em By (\ref{3.1}) and the current conservation
law $ \pa_{\mu}q^{\mu}_k=0$, we have the speed of $e_k$
\begin{equation}
{\vec v}_k=\int\vec{x}\pa_0 q_k^0d^3x=-\int\vec{x}(\nb\cdot
\vec{q})d^3 x=\int \vec{q}_kd^3 x. \label{3.2}\end{equation}}

{\bf Lemma 2.} {\em For any Hermitian operator $\hat P$, and
corresponding classical quantity $P_k$ for $e_k$ by
\begin{equation}
P_k\equiv \int_{R^3}\phi_k^+\hat P\phi_k d^3x,
\end{equation}
then we have
\begin{equation}
 \frac d {dt} P_k=\int_{R^3}
\phi_k^+\left(\pa_t \hat P+\frac i \hbar[\hat H_k,\hat
P]\right)\phi_k d^3 x,\label{poss}
\end{equation}
where $[\hat H_k,\hat P]=\hat H_k\hat P-\hat P\hat H_k$. }

{\bf Theorem 1.} {\em For 4-d momentum $p^\mu_k$ of each $e_k$, we
have the following rigorous dynamic equation
\begin{equation}
\left\{
\begin{array}{l} \frac d {dt}p^0_k =\int\vec q_k\cdot(e\vec
E-w\vec{\mathcal{E}}_k)d^3 x-\frac d{dt} W_k  ,\\
\frac d {dt}\vec p_k =\int [q^0_k(e \vec E-w \vec{\mathcal
E}_k)+\vec q_k\times(e\vec B-w\vec {\mathcal B}_k)] d^3 x,
\end{array}\right. \label{dnm} \end{equation}
where $\vec E$ and $\vec B$ are electrical and magnetic
intensities given in (\ref{2.8}), $\vec{\mathcal{E}}_k$ and
$\vec{\mathcal{B}}_k$ are intensities caused by current except
$e_k$ its own current
\begin{equation}
\left\{
\begin{array}{l} \vec{\mathcal{E}}_k =-(\nb Q^0+\pa_0\vec{Q})+(\nb q_k^0+\pa_0\vec{q_k})
=-\sum_{l\neq k}(\nb q_l^0+\pa_0\vec{q_l}),\\
\vec{\mathcal{B}}_k = \nb \times \vec {Q}-\nb \times \vec
{q_k}=\sum_{l\neq k}\nb \times \vec {q_l}.
\end{array}\right. \label{weak} \end{equation}}

{\bf Proof.} For (\ref{3.3}), we have momentum and energy operator
of $e_k$
\begin{equation}\left\{
\begin{array}{l} \hat p^0_k
=\hat H_k-eA^0+w(Q^0-q^0_k)=\vec \al\cdot\hat p_k+\mu
\ga- \ck\ga_k\ga, \\
\hat{p}_k=-\hbar i\nb-e\vec A+w(\vec Q-\vec q_k),~~\hat K_k=\hat
p_k^0+\frac 1 2 w \ck\ga_k\ga.\end{array}\right.
\label{pkv}\end{equation} Then by straightforward calculation we
have
\begin{equation}
\left\{
\begin{array}{l}
\pa_t \hat K_k = \vec \al\cdot \pa_0  [-e \vec A+w(\vec Q-\vec q
_k)]-\frac 1 2 w\pa_0\ck\ga_k \ga, \\
\pa_t \hat p_k=\pa_0  [-e \vec A+w(\vec Q-\vec q _k)],
\end{array}\right.
\label{22}\end{equation}
\begin{equation}
[\hat H_k, \hat K_k] = \vec \al \cdot(-\hbar
i\nb)[-eA^0+w(Q^0-q^0_k)]+ \frac 1 2 w (\hat
H_k\ck\ga_k\ga-\ck\ga_k\ga\hat H_k),\label{23}\end{equation}
\begin{equation}\begin{array}{lll}
[\hat H_k, \hat p_k]&=&[eA_0-w(Q_0-q_{k0})+(\mu-w
\ck\ga_k)\ga+\vec\al\cdot\hat p_k~,~\hat p_k]\\
&=&\hbar i\nb[eA_0-w(Q_0-q_{k0})]+ w \hbar i\nb
\ck\ga_k\ga-\vec\al\times(\hat p_k\times\hat p_k)\\
&=&\hbar i\nb[eA_0-w(Q_0-q_{k0})]+ w \hbar i\nb
\ck\ga_k\ga-\\&~&\hbar i\vec\al\times\{\nb\times[e\vec A-w(\vec
Q-\vec q_k)]\}.
\end{array}\label{30}\end{equation}
So we can derive
\begin{equation}
\begin{array}{lll}
\frac d {dt}K_k &=&\int\phi_k^+[\pa_0 \hat K_k +\frac i \hbar(
\hat H_k \hat K_k-\hat K_k\hat H_k)]\phi_k d^3x\\
&=&\int\phi_k^+ \vec \al \cdot\left(\pa_0 [-e \vec A+w(\vec Q-\vec
q _k)]+\nb[-eA^0+w(Q^0-q^0_k)]\right)\phi_k d^3 x+\\
&~&\frac 1 2 w \int\phi_k^+\left(-\pa_0\ck\ga_k \ga+ \frac i \hbar
(\hat H_k\ck\ga_k\ga-\ck\ga_k\ga\hat H_k)\right)\phi_k d^3 x\\
&=&\int\left(\phi_k^+\vec \al \cdot(e\vec
E-w\vec{\mathcal{E}}_k)\phi_k -\frac 1 4 w\pa_0\ck\ga_k^2+\frac i
{2\hbar}w\phi_k^+(\hat H_k\ck\ga_k\ga-\ck\ga_k\ga\hat
H_k)\phi_k\right) d^3 x \\
&=&\int\vec q_k\cdot(e\vec E-w\vec{\mathcal{E}}_k)d^3 x-\frac 1 4w
\int\pa_0\ck\ga_k^2d^3 x+\\
&~& \frac i {2\hbar}w \int[(\hat
H_k\phi_k)^+\ck\ga_k\ga\phi_k-\phi_k^+\ck\ga_k\ga(\hat H_k\phi_k)]
d^3 x\\
&=&\int\left(\vec q_k\cdot(e\vec E-w\vec{\mathcal{E}}_k)-\frac 1 4
w \pa_0\ck\ga_k^2+\frac 1 2 w
\ck\ga_k[(\pa_0\phi_k)^+\ga\phi_k+\phi_k^+\ga(\pa_0\phi_k)]
\right)d^3 x\\
&=& \int\vec q_k\cdot(e\vec E-w\vec{\mathcal{E}}_k)d^3 x.
\end{array} \label{24} \end{equation}
\begin{equation}
\begin{array}{lll} \frac d {dt}\vec p_k
&=&\int\phi_k^+[\pa_0 \hat p_k +\frac i \hbar(
\hat H_k \hat p_k-\hat p_k\hat H_k)]\phi_k d^3x\\
&=&\int q^0_k\left(\pa_0  [-e \vec A+w(\vec Q-\vec q
_k)]-\nb[eA_0-w(Q_0-q_{k0})]\right)d^3 x-\\&~& w\int \nb
\ck\ga_k^2 d^3 x+\int \vec q_k\times\left(\nb\times[e\vec A-w(\vec
Q-\vec q_k)]\right)
d^3 x\\
&=&\int [q^0_k(e \vec E-w \vec{\mathcal E}_k)+\vec q_k\times(e\vec
B-w\vec {\mathcal B}_k)] d^3 x
\end{array}\label{25}\end{equation}
The proof is finished by $\hat p_k^0=K_k-W_k$.

Form the calculation of (\ref{24}) and (\ref{25}) we learn that
the forces caused by vector current have the same form as that
caused by electromagnetic field. Considering that the
corresponding force is much less than that of electromagnetic
field, we omit $\mathcal{\vec E}$ and $\mathcal{\vec B}$ in the
following analysis.

\section{The Classical Approximation}
\setcounter{equation}{0}

Assume the external potential is adequately small and the
distances among electrons are adequately long, then all
$\phi_k(t)$ are solitary waves with central coordinate ${\vec X}_k
(t)$. By (\ref{2.5}), (\ref{3.1}) and (\ref{3.2}), we get
\begin{equation}
q_{k0}\to \dl ({\vec x}-{\vec X}_k ), \quad {\vec q}_k\to \dl
({\vec x}-{\vec X}_k ){\vec v}_k.
 \label{3.8}\end{equation}
\begin{equation}
W_k=\sqrt{1-v^2_k}W,\quad W\equiv \frac 1 2 w \int \ck\ga_k^2
d^3\bar x_k,
 \label{38}\end{equation}
where $\bar x_k$ stands for the central reference coordinate
system of $e_k$, and $W$ is the proper energy corresponding to
nonlinear term. Substituting (\ref{3.8}) into (\ref{dnm}), we get
the Newton's second law for all electrons as follows
\begin{equation} \left
\{\begin{array}{l} \frac d {dt}p_k^0=e{\vec v}_k\cdot{\vec
E}({\vec
X}_k )-\frac d {dt} W_k, \\
\frac d {dt}{\vec p}_k=e({\vec E}+{\vec v}_k\times {\vec B}).
\end{array} \right.
 \label{3.9}\end{equation}
The proper time of $e_k$ reads $d\tau_k\equiv\sqrt{1-v_k^2}d t$,
the $4-d$ speed $
 u^{\mu}_k=(1,{\vec v}_k )/\sqrt{1-v_k^2}$.
Correspondingly, the potential equation becomes
\begin{equation}
\pa_\nu\pa^\nu A^{\mu}=e\sum^N_{k=1}v_k^{\mu} \dl({\vec x}-{\vec
X}_k )+\rho^{\mu},\quad v^{\mu}_k\equiv (1,{\vec v}_k ).
 \label{3.12}\end{equation}

In what follows, we look for the relation between $p^\mu_k$ and
$u^\mu_k$, and the definition of inertial mass $m_e$. By
(\ref{3.9}) we get
\begin{equation}
u^{\mu}_k \frac d {dt}p_{k\mu}=-u^0\frac d{dt}W _k= -\frac
W{\sqrt{1-v^2_k}} \frac d{dt}\sqrt{1-v_k^2}.
\label{3.13}\end{equation} By (\ref{2.9}) and the Pauli-Fierz
identity, we rewrite (\ref{2.2}) as
$$ [-\hbar i \nb+e{\vec
A}-w({\vec Q}-{\vec q}_k)]\phi_k={\vec \al}[\hbar
i\pa_0-eA_0+w(Q_0-q_{k0})]\phi_k+\vec S\phi_k,
$$
where ${\vec S}\equiv (S_1,S_2,S_3)$ are all anti-Hermitian
matrices. By the above equation and (\ref{3.3}), we have
$$
 p^\mu_k=\frac{1}{2}\int_{R^3}[(\hbar i\pa_0\phi_k)^{+}\al^\mu\phi_k+\phi^{+}_k \al^\mu(\hbar i
\pa_0\phi_k)]d^3x-\int_{R^3}[eA_0-w(Q_0-q_{k0})]q^\mu_k d^3x.$$ If
$e_k$ moves in a infinitesimal speed ${\vec v}_k $, then $\hbar
i\pa_0\phi_k \to \bar m \phi_k$, where $\bar m>0$ is a constant.
Hence by (\ref{3.8}) we get
\begin{equation}
p^\mu_k= \int[\bar m -eA_0+w(Q_0-q_{k0})] q^\mu_k d^3x \equiv m_k
u^\mu_k , \label{3.15}\end{equation}
where $m_k=\sqrt{p^\mu_k
p_{k\mu}}$ is the scalar mass of $e_k$. $m_k$ times (\ref{3.13})
we have
$$\frac 1 2 \frac{d}{d t}(p^{\mu}_k p_{k\mu}) = -m_k
W\frac d {dt} {\rm ln}\sqrt{1-v_k^2},{\rm~~~or~~~} \frac{d}{d
t}m_k = - W\frac d {dt} {\rm ln}\sqrt{1-v_k^2},\nonumber  $$ so we
can define the inertial mass of electron $e_k$ as
\begin{equation}
m_k=m_e+W{\rm ln}\frac 1 {\sqrt{1-v^2_k}},
\label{3.16}\end{equation} where $m_e$ is static mass of an
electron. Substituting (\ref{3.16}) into (\ref{3.15}) and
(\ref{3.3}) we get
\begin{equation}\left\{\begin{array}{l}
p^{\mu}_k=\left(m_e+W{\rm ln}\frac 1 {\sqrt{1-v^2_k}}\right)u^{\mu}_k\\
K_k=\frac{m_e}{\sqrt{1-v^2_k}}+W\left(\frac 1 {\sqrt{1-v^2_k}}
{\rm ln} \frac 1
{\sqrt{1-v^2_k}}+{\sqrt{1-v^2_k}}\right)\end{array}\right.
 \label{3.17}\end{equation}
where $m_k$ is the moving inertial mass of $e_k$. From
(\ref{3.17}) we find that the nonlinear term violates mass-energy
relation slightly. Although (\ref{3.17}) is derived under the
assumption of infinitesimal speed, it is also suitable for the
case of high speed, because of the covariant form of $p^\mu_k$.

Since $W\ll m_e$, we omit it from (\ref{3.17}). Then the
Lagrangian corresponding to the coupling system (\ref{3.9}) and
(\ref{3.12}) reads
\begin{equation}
{\cal L}=\sum^N_{k=1}(-m_e-eA_{\mu}u^{\mu}_k)
 \sqrt{1-v_k^2}\dl({\vec x}-{\vec X}_k)+eA_{\mu}\rho^{\mu}-\frac{1}{2}\pa_{\mu}A_{\nu}\pa^{\mu}A^{\nu}.
 \label{3.18}\end{equation}

If all $|\vec v_k|\ll 1$ and the effect of the retarded potential
can be ignored, then the nonrelativistic approximation of the
total potential becomes
\begin{equation}
A^{\mu}=V^{\mu}({\vec x})+\frac{e}{4\pi
}\sum^N_{k=1}\frac{v^{\mu}_k} {|{\vec x}-{\vec X}_k|}.
\label{3.22}\end{equation} Omitting the self-potential in
(\ref{3.18}) we get an ordinary differential system
$$
{\cal L}_e=-\sum^N_{k=1}\left(m_e\sqrt{1-v_k^2}+e
V_{\mu}v_k^{\mu}+\frac{e^2} {8\pi }\sum_{l\ne
k}\frac{v_{l\mu}v^{\mu}_k}{|{\vec x}-{\vec X}_l|}\right) \dl({\vec
x}-{\vec X}_k),
$$
\begin{equation}
L=\int_{R^3}{\cal L}_ed^3x=-\sum^N_{k=1}\left(m_e\sqrt{1-v_k^2}+e
V_{\mu}(\vec X_k)v_k^{\mu}+\frac{e^2} {8\pi }\sum_{l\ne
k}\frac{v_{l\mu}v^{\mu}_k}{|{\vec X_k}-{\vec
X}_l|}\right).\label{3.23}
\end{equation}
The Hamiltonian of the system reads
\begin{equation}
H=\sum_{k=1}^N \frac {\pa L}{\pa \vec v_k}\cdot\vec
v_k-L=\sum^N_{k=1}\left(\frac {m_e}{\sqrt{1-v_k^2}}+e
V_{k0}+\frac{e^2} {8\pi }\sum_{l\ne k}\frac1{|{\vec X_k}-{\vec
X}_l|}\right).\label{3.14}
\end{equation}

(\ref{3.18}), (\ref{3.23}) and (\ref{3.14}) constitute the
complete theory of the classical mechanics for $N$-electron.

\section{The Nonrelativistic Approximation}
\setcounter{equation}{0}

Now we make the nonrelativistic approximation and derive the
Schr\"odinger equation for $N$-electron which move slowly in
strong external potential. The conventional quantum mechanics is a
linear theory, which defaults the following hypotheses:

 {\bf Hq.1} The effects of retarded potential can be ignored.

 {\bf Hq.2} All nonlinear coupling terms can be ignored.

 {\bf Hq.3} All self coupling potentials, including electromagnetic
 field caused by electron its own, can be merged into physical
 mass of electron.

In what follows, we also accept {\bf Hq.1-3} as auxiliary
assumptions adding upon equation (\ref{2.2}) and (\ref{2.8}). The
{\bf Hq.1} means that, only low speeds $(v_k\ll 1)$ are considered
and the distances among electrons are adequately small. The {\bf
Hq.2} means that, the potential caused by electrons is much less
than the external potential $(|A^\mu_k|\ll |V^\mu|)$, so that the
second order terms can be omitted. The {\bf Hq.3} means that the
total effect of self-coupling potentials is very small $(\ll \mu)$
and stable, so it can be merged into the physical mass of an
electron $m$. The following procedure was once used to derive the
Breit potential among electrons with $O(\frac 1 {m^2})$
terms\cite{30}. Here we only keep $O(\frac 1 {m})$ terms to
demonstrate how (\ref{2.2}) implies the $N$-body Shr\"odinger
equation.

Denote $A_k^{\mu}$ is potential caused by $e_k$, i.e.,
$\pa_\al\pa^\al A_k^{\mu}=e q_k^{\mu}$.  By {\bf Hq.1} we have
solution without retarded potential as follow
\begin{equation}
A_k^{\mu}=\frac{e}{4\pi}\int_{R^3} \frac{q^{\mu}_k (t,\vec X_k)}
 {|\vec{x}-\vec{X}_k|} d^3 X_k.
 \label{4.1}\end{equation}
By {\bf Hq.2} and {\bf Hq.3}, (\ref{2.2}) becomes
\begin{equation}
(\hbar i\pa_0-e\Phi_{k0})\phi_k=\vec{\al}\cdot (-\hbar i\nb+e\vec
{\Phi}_k)\phi_k+m\ga\phi_k, \label{4.2}\end{equation}  where
$\Phi^{\mu}_k=A^{\mu}-A_k^{\mu}$ is the external potential with
respect to $e_k$.

Taking conventional transformation
\begin{equation}
\phi_k={\rm exp}\left(\frac{m c^2t}{\hbar
i}\right)\left(\begin{array}{l} \psi_k
\\ \dl_k \end{array} \right). \label{4.3}\end{equation}
where $\psi_k$ and $\dl_k$ are slowly varying functions of $t$.
Substituting (\ref{4.3}) into (\ref{4.2}) and omitting $O(\frac 1
{m^2})$ we get
\begin{equation}
\dl_k=\frac 1{2m}\vec{\sg}\cdot (-\hbar i
 \nb+e\vec{\Phi}_k)\psi_k,
 \label{4.4}\end{equation}
\begin{equation}
\hbar i\pa_0\psi_k=e\Phi_{k0}\psi_k+\vec{\sg}\cdot (-\hbar i \nb
+e\vec{\Phi}_k)\dl_k. \label{4.5}\end{equation}  Substituting
(\ref{4.4}) into (\ref{4.5}) we get the Pauli's equation for each
$e_k$
\begin{equation}
 \hbar i \pa_t\psi_k = \left(-\frac{{\hbar}^2}{2m}\Delta+e\Phi_{k0}+\mu_z\vec{\sg}\cdot
  \vec{B}+\frac{e}{m}\vec{\Phi}_k\cdot (-\hbar i
\nb)+\frac{e^2}{2m}\vec{\Phi}_k^2\right)\psi_k.
\label{4.6}\end{equation} where $\mu_z=\frac{|e|\hbar}{2m}$ is the
Bohr magneton. The Coulomb gauge $\nb\cdot \vec{\Phi}_k=0$ holds
in (\ref{4.6}) by {\bf Hq.1}.

Since the magnetic field caused by $\vec{A}_k$ is much smaller
than that of external potential, we also omit it for simplicity.
Let $x_3$ be parallel with $\vec{B}$, then
$$\vec{\sg}_k\cdot\vec{B}=\sg_3 B=\la_k B,\quad
\la_k=1~ {\rm or} ~-1.$$
Hence one component of $\psi_k$ vanishes,
so all $\psi_k$ can be treated as a scalar. By {\bf Hq.2},
(\ref{4.6}) becomes
\begin{equation}
\hbar i \pa_t\psi_k=\left[\frac 1 {2m} \hat P^2+eV_0+\la_k\mu_zB+
e\sum_{l\ne k}\left(A_{l0}+\frac{1}{m}\vec{A}_l\cdot\hat P
\right)\right]\psi_k,
 \label{4.7}\end{equation}
where $\hat P=-\hbar i\nb+e{\vec V}$. Substituting (\ref{4.4})
into (\ref{2.3}) we get
\begin{equation}
q_{k0}=|\psi_k|^2,\quad
\vec{q}_k=\frac{1}{2m}[\psi^{+}_k\hat{P}\psi_k+ (\hat
P\psi_k)^{+}\psi_k] .
 \label{4.9}\end{equation}
By (\ref{4.1}) and {\bf Hq.1} we get
\begin{equation}
A_{k0}=\frac{e}{4\pi }\int_{R^3}\frac{|\psi_k(t,\vec X_k)|^2}
{|\vec{x}-\vec{X}_k|}d^3 X_k, ~ \vec{A}_k=\frac{e}{4\pi}\int_{R^3}
\frac{\psi^{+}_k{\hat P}_k\psi_k}{m|\vec{x}-\vec{X}_k|} d^3 X_k,
 \label{4.10}\end{equation}
where $${\hat P}_k=-\hbar i\nb_k+\vec{V}(t,{\vec X}_k)=-\hbar
i\frac {\pa}{\pa\vec X_k}+\vec{V}(t,{\vec X}_k).$$

Substituting (\ref{4.10}) into (\ref{4.7}) we get the separating
form for nonrelativistic quantum mechanics for $N$-electron as
follows
\begin{equation}
\hbar i \pa_t \psi_k=\left(\frac 1{2m}{\hat
P^2}+eV_0+\la_k\mu_zB+\frac{e^2}{4\pi}\sum_{l\ne k}\int_{R^3}
\frac{|\psi_l({t,\vec X}_l)|^2}{|\vec{x}-\vec{X}_l|}d^3 X_l
\right)\psi_k,\label{4.11}\end{equation}  The action corresponds
to (\ref{4.11}) is given by
\begin{equation}\begin{array}{lll}
 {\textbf{I}}&=&\sum^n_{k=1} \int^t_{t_0}d t\int \psi^{+}_k(t,\vec{x})(\hbar i
 \pa_t-{\hat\textbf{H}_k} )\psi_k(t,\vec x)d^3x\\
 &=&\sum^n_{k=1} \int^t_{t_0}d t\int \psi^{+}_k(t,\vec{X}_k)(\hbar i
 \pa_t-{\hat\textbf{H}_k} )\psi_k(t,\vec {X}_k)d^3{X}_k,\end{array}
 \label{4.13} \end{equation}
where all coordinates are integral variables, i.e., they are dummy
arguments, so we can assign $\vec{X}_k$ to the electron $e_k$,
\begin{equation}{\hat\textbf{H}_k}=\frac 1{2m}{\hat
P^2_k}+eV_0(t,\vec X_k)+\la_k\mu_zB(t,\vec
X_k)+\frac{e^2}{8\pi}\sum_{l\ne k}\int_{R^3} \frac{|\psi_l({t,\vec
X}_l)|^2}{|\vec{X}_k-\vec{X}_l|}d^3 X_l.\end{equation}

Noticing the normalizing conditions
\begin{equation}
1=\int_{R^3}|\psi_k(t,\vec X_k)|^2 d^3 X_k, \quad
(k=1,\cdots,N),\label{4.14}
\end{equation}
Multiplying (\ref{4.13}) by (\ref{4.14}), we get a combined form
of action
\begin{equation}
 {\textbf{I}}=\int^t_{t_0}d t \int_{R^{3n}}\Psi^{+}(\hbar i
\pa_t-\hat{{\textbf{H}}})\Psi d^3 X_1 d^3 X_2\cdots d^3 X_n,
 \label{4.15}\end{equation}
where
\begin{equation}
\Psi(t,\vec{X}_1,\vec{X}_2, \cdots, \vec{X}_n)=\psi_1(t,\vec{X}_1)
 \psi_2 (t,\vec{X}_2)\cdots \psi_n (t,\vec{X}_n),
 \label{4.16} \end{equation}
the total Hamiltonian operator is given by
\begin{equation}
\hat {\textbf{H}}=\sum^n_{k=1}\left(\frac{1}{2m}{\hat
P}_k^2+eV_0(t,\vec X_k) +\la_k\mu_z B(t,\vec X_k)+\frac{e^2}{8\pi
}\sum_{l\ne k}\frac{1}
 {|\vec{X}_l-\vec{X}_k|}\right).
 \label{4.17}\end{equation}
By variation of (\ref{4.15}) with respect to $\Psi$, we finally
get the standard Schr\"odinger equation for $N$-electron
\begin{equation}
\hbar i \pa_t \Psi=\hat {\textbf{H}} \Psi.  \label{4.19}
\end{equation}
Comparing (\ref{4.17}) with (\ref{3.14}), we find that the
Hamiltonian of the classical mechanics and quantum mechanics have
just the same structure, except a magneton term which is ignored
in classical approximation.

\section{Discussion and Conclusion}
The above derivation clearly shows how a self consistent system
(\ref{2.2}) naturally implies the classical and quantum mechanics
for many-body. But the nonlinear coupling potential in (\ref{20})
is not finally determined. What is more important for this work is
that, it shows us there are different solving methods for a
specific physical problem.

In the procedure of transition from the classical mechanics to the
quantum theory, the correspondence principle is used, i.e., the
classical quantities are translated into the corresponding
operators in quantum mechanics by analogy. From the above
derivation we find that the analogies are fortunately valid for
some cases. The Schr\"odinger equation is a typical example of
success. But analogy is not reliable generally, we should keep
some suspicion and caution in mind while doing so, rather than
treat it absolutely.

The procedure of second quantization seems to expand one Dirac's
equation into many ones by applying Fourier transform and the
principle of superposition. It seems to be closely related with
the dynamic equation (\ref{2.2}). However this procedure is mainly
effective for linear equation, so it must continuously introduce
extra conditions such as the principle of exclusion, positive
energy condition, quantum condition, box normalizing condition,
renormalizing procedure, etc. as compensations to keep the quantum
theory consistent.

Despite each approach to the many body quantum mechanics starting
from the different points, it is surprising that all approaches
have achieved successes to some degree for several cases. There
should be some intrinsic relations between these theories, in
which the action principle seems to play a central role and act as
a connecting bridge.

\section*{Acknowledgments}
I am grateful to Prof. Jaime Keller, Prof. Guang-Jiong Ni and my
friends Prof. De-Xing Kong, Prof. Hao Wang, Prof. Er-Qiang Chen
and Miss Sunny Sun for their encouragement and kind help.

\end{document}